\documentclass[twocolumn,secnumarabic,amssymb,nobibnotes,aps,prd]{revtex4-1}
\usepackage[T1]{fontenc}
\usepackage{graphicx}
\usepackage{dcolumn}
\usepackage{bm}
\usepackage{float}
\usepackage{sidecap}
\usepackage{color}
\usepackage{epstopdf}
\graphicspath{{Pictures/}}

\begin{document}

\title{Detection of the adsorption of water monolayers through the ion oscillation frequency in the magnesium oxide lattice by means of Low Energy Electron Diffraction}
  
\author{M. Guevara-Bertsch$^{1,2}$, G. Ram\'irez-Hidalgo$^{2,3}$, A. Chavarr\'ia-Sibaja$^{2}$, E. Avenda\~{n}o$^{1,2}$, J.A. Araya-Pochet$^{2}$, and O.A. Herrera-Sancho$^{1,2,4}$}

\address{$^1$ Escuela de F\'isica, Universidad de Costa Rica, 2060 San Pedro, San Jos\'e, Costa Rica}
\address{$^2$ Centro de Investigaci\'on en Ciencia e Ingenier\'ia de Materiales, Universidad de Costa Rica, 2060 San Pedro, San Jos\'e, Costa Rica}
\address{$^3$ Secci\'on de F\'isica Te\'orica, Universidad de Costa Rica, 2060 San Pedro, San Jos\'e, Costa Rica}
\address{$^4$ Institut f\"ur Quantenoptik und Quanteninformation,\"Osterreichische Akademie der Wissenschaften, Technikerstr. 21a, 6020 Innsbruck, Austria}
\date{\today}%
\vspace{10pt}

\begin{abstract}

We investigate the variation of the oscillation frequency of the Mg$^{2+}$ and O$^{2-}$ ions in the magnesium oxide lattice due to the interactions of the surface with water monolayers by means of Low Energy Electron Diffraction. Our key result is a new technique to determine the adsorbate vibrations produced by the water monolayers on the surface lattice as a consequence of their change in the surface Debye temperature and its chemical shift. The latter was systematically investigated for different annealing times and for a constant external thermal perturbation in the range of 110--300\,K in order to accomplish adsorption or desorption of water monolayers in the surface lattice.\\   
\smallskip
\noindent \textbf{PACS number(s):} 61.05.jh, 68.43.Pq, 82.65.+r

\end{abstract}

\maketitle
\section{Introduction}

\textit{``God made the bulk; the surface was invented by the devil''}, Noble Prize Wolfgang Pauli used to say as he referred to the exciting challenge of studying the first monolayers of a sample~\cite{Grote:2011}. He explained  that the complex properties of surfaces were due to surface atoms not having an isotropic environment; they interact with three different types of atoms: those in the bulk below, those from the same surface, and those in the neighborhood. This paper focuses on the study of the interactions of the surface of a magnesium oxide crystal with the environment through the measurement of the variation of the oscillation frequency of the lattice by means of Low Energy Electron Diffraction (LEED). 

Magnesium oxide (MgO) is commonly used in surface science studies, due to its high Poisson'\,s ratio and low Gibb'\,s free energy. Furthermore, it is used as a lattice template for growing oriented ferroelectric and superconducting nanostructures, which has led to an increasing number of applications. For this reason, understanding the procedures needed to prepare MgO surfaces of very high quality is of higher importance in numerous areas of surface science. As reported in Refs.~\cite{Moses:2010,Cookt:1982}, an increasing number of researchers have been engaged in the fabrication and characterization of MgO crystals. It is noteworthy that MgO is an unique solid due to its highly ionic character, simple stoichiometry and rock--salt crystal structure, single valence state, and only one stable low--index, $\left\{100\right\}$, surface orientation~\cite{Heinrich:1994}. 

The interaction of water with MgO is one of the central points in the characterization of these substrates. As explained in Ref.~\cite{Gaddy:2014}, the understanding of the reaction of water vapor with metal oxides is the key to creating multifunctional devices such as quantum--well superstructures, high--mobility transistors and laser diodes, among other applications. MgO has a simple face-centered-cubic (fcc) structure and an absence of \textit{d}--orbital electrons, which makes it an ideal model system for the study of these interactions. Over the last years diverse techniques have been applied to characterize the adsorption processes of water films on the surface of MgO substrates as reported in Ref.~\cite{Coulomb:2013}. It is needless to say that the development of a clear understanding of the chemistry of water--solid interfaces is essential in many phenomena within environmental protection, geology, atmospheric chemistry, archeology, corrosion, sensors, heterogeneous catalysis, and electronics~\cite{Freund:1996}.

Low Energy Electron Diffraction has become the prime technique used to determine atomic locations at surfaces, as explained in Ref.~\cite{Hove:1986}. Due to its low energy range 10--1000\,eV, LEED is structurally sensitive in variations to the order of 5--10\,{\AA}. Any variation in the oscillation frequency of the atoms in the lattice can be observed through the variation of the intensity of the LEED pattern. It is well known that LEED intensities depend on temperature, i.e. on the thermal vibrations of the atoms on the surface of the crystal. Therefore, it is possible to measure the Debye temperature of solid surfaces, which is related to the dynamic motions of atoms and it is vital to the understanding of the new physical properties arising when the translational symmetry of a three-dimensional solid is broken~\cite{Brune:2009}. 

This article reports on an investigation of the determination of the surface Debye temperature by means of LEED of the MgO(100) crystal, in order to understand the process of adsorption (or desorption) of water molecules on the crystal surface for the temperature range of 110--300\,K. We also determine the chemical shift of the O--2p shell induced by the interactions of the water monolayers for different initial conditions of the surface lattice. Through this approach we can estimate the number of water monolayers that can be adsorbed or desorbed from the MgO surface at different annealing times.  

This paper is organized as follows: the experimental setup is described in Section~\ref{section:exp}. The processing techniques applied to obtain the variation of intensity of the LEED patterns in the MgO crystal are described in Section~\ref{section:video}. In Section~\ref{section:patterns} this data is used to obtain the dependence of the intensity of the LEED patterns with temperature. Using the above information the surface Debye temperature is determined for different physical conditions in Section~\ref{section:debye}.

\section{Experimental setup} \label{section:exp}
We use two different commercially available MgO crystals with a (100) crystallographic plane orientation. The dimensions of the single crystals are 10\,mm$\times$10\,mm$\times$0.5\,mm and one set has two epi-polished sides while the other one has only one polished side. The purity of the crystals is higher than 99.9\% and the stated principal impurities present in the crystal are calcium, aluminum, silicon, iron, chromium, boron and carbon in a level lower than 50\,ppm~\cite{Harris:2013}. Before our LEED studies of the crystal surface, X-ray diffraction patterns were obtained by a Bruker (model: D8 Advance) diffractometer with Cu-K$_{\alpha1}$ ($\lambda$=0.154\,0562\,nm) and Cu-K$_{\alpha2}$ ($\lambda$=0.154\,4398\,nm) emission lines as the radiation source and a LynxEye detector.

\begin{figure}
\begin{center}
\includegraphics[height=6cm]{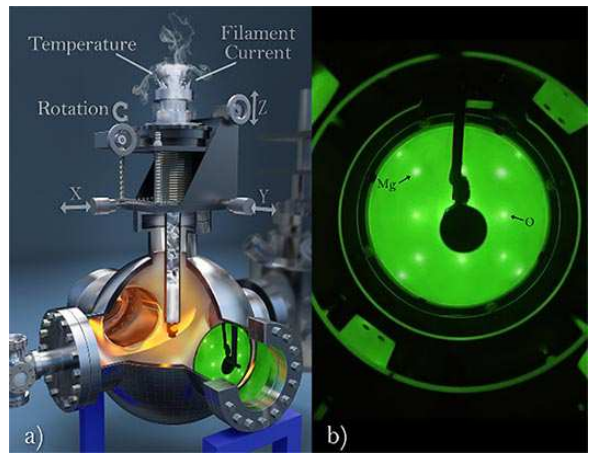}
\caption{(Color online) Left panel (a): artistic visualization of our experimental set up. Translational and rotational degrees of freedom for the MgO crystals are shown with arrows. Two multipin electrical feedthroughs are used to measure the temperature and apply a current to the filament. The middle bar that holds the MgO crystals could be heated or cooled by means of electrons from a filament and liquid nitrogen, respectively. Right panel (b): LEED pattern of the fcc MgO lattice showing the positions of the Mg$^{2+}$ and O$^{2-}$ ions. As shown with arrows, the O$^{2-}$ ions occupy the octahedral fcc lattice position and the Mg$^{2+}$ ions occupy the octahedral holes.}
\label{experimental}
\end{center}
\end{figure}

The crystals were mounted inside an ultra--high vacuum apparatus, modified from that of Ref.~\cite{Herrera:2011}, at the pressure of $\approx$\,10$^{-9}$\,Torr, as seen in Fig.~\ref{experimental}(a). By means of a tungsten filament and a liquid nitrogen reservoir, it is possible to change the crystal temperature from 110\,K to 820\,K. A copper-constantan thermocouple is attached to the crystal holder in order to keep a precise temperature control during the measurements. The LEED patterns were obtained in situ with a commercial ErLEED 1000-A from the SPECS company at the vacuum atmospheres described above. The acceleration voltage of the electron gun ranges from 0\,eV to 1000\,eV, and typically the screen voltage was held about 4.5\,keV, the filament current was around 2.5\,A, and the emission current was approximately 300\,$\mu$A. The LEED patterns were externally recorded with a camera attached to the outside window of the phosphorous screen and the intensity, size, and position of the diffraction spots were analyzed later as described below in Section~\ref{section:video}. A typical LEED pattern for the MgO crystals is shown in Fig.~\ref{experimental}(b), where the crystallographic array for magnesium and oxygen ions are shown~\cite{Ferry:1996}.   
            
\section{Data processing}\label{section:video}

As explained in Ref.~\cite{Hove:1986} there are several different types of LEED intensity measurements which may be made, depending on the purpose of the experiment. It is possible, for example, to measure the intensity versus energy or accelerating voltage (I-V curves) of the incident electron beam for the study of the surface structure. Other examples are: the intensity versus polar angle of emergence (I-e curves) for the analysis of the instrumental response, and intensity versus temperature of the crystal surface (I-T curves) for determination of the Debye temperature of the crystal. In order to measure the intensity of the diffraction spots from the LEED patterns, a computer based program using Python and Bash scripting was developed. Due to the large amount of data obtained in the experiment it became necessary to develop a more efficient method to process the information in real time. We also wanted to have a better control of the precision and resolution of the measured parameters, for example, on the reduction of the background effects. These two features where difficult to obtain in commercially available programs. In a simple manner, the program quantifies the intensity of the spots in each frame of a video by counting the number of pixels in the regions determined by the user. 

In the first stage, the user draws circles around the diffraction spots as start appearing in the video, as shown in Fig.~\ref{experimental}(b). Once all the regions of study are specified, the program measures the amount of pixels in each circle for every frame in the video. Thereupon the program retrieves the information of the position ("\textit{x}" and "\textit{y}" coordinates) of the studied spots and the average intensity for every spot in each frame. For a better visualization and analysis, the information is saved (in a specific directory) in two ways: (1) as a 2D graph of the position of the spots, and (2) as a plot of the variation of the intensity of the diffracted spots against time. This information is generated for each circle specified by the user. 

The program offers the versatility to process the videos in two different modes according to the experiment, namely: Full-- and Segmented--mode. The Full--mode processes a complete video specified by the user without any modification. This mode is applied when the experiment requires the analysis of only one video, for example to measure the variation of the intensity as a function of the acceleration voltage (I-V curves). The Segmented--mode is used when it is necessary to process a set of different videos, for example to determine the variation of the intensity at different temperatures (I-T curves). In this case, a different video of the same length is taken for each temperature. The program takes one representative frame per video and then merges all of them. Continuing in the Segmented--mode, the representative extraction is done in two ways: Middle-- and Average--mode. In the Middle--mode the representative frame is simply the middle frame of each video and the Average--mode is the average intensity for every pixel for the whole set of frames of the video. Finally the resulting video is analyzed in the Full--mode. 

To reduce the background effects from the data, the program quantifies the intensity from a ring surrounding the drawn area and then subtracts this value from the intensity of the spot. The entire program has been developed with free and open computational tools, and tested in comparison with the commercially available software Icy\textsuperscript{TM}. The results are in accordance with the ones obtained from the software with the exception that execution time is considerably lower (approximately 80\% less) and the background is subtracted in a systematic manner. Time execution, in our program, is around 10\,min and  depends mainly on one factor: how many contours are drawn for the whole video. The program can be accessed at \url{https://github.com/Gustavroot/LEED_CICIMA} and is open to modifications and improvements.

\section{Variation of LEED patterns with temperature (I-T curves)} \label{section:patterns}

\begin{figure}[H]
\begin{center}
\includegraphics[height=5cm]{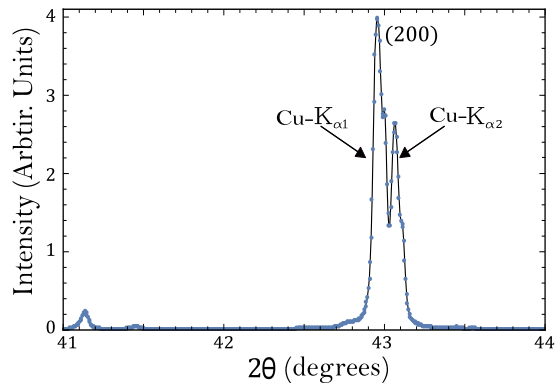}
\caption{(Color online) Coherent elastic scattering of X-rays in the MgO crystal. The double peaks at (200) corresponds to the Cu-K$_{\alpha1}$ and Cu-K$_{\alpha2}$ transition energies of the X-ray source. See the text for further details.}
\label{xray}
\end{center}
 \end{figure}

Before applying the ultra--high vacuum atmospheres, X-ray diffraction patterns of the MgO crystals were measured to obtain crystallographic information. A typical X-ray diffraction pattern is shown in Fig.~\ref{xray} with the crystal orientation $(hkl)$, as labeled. The crystal orientation obtained in our measurements agrees with Ref.~\cite{Stampe:1998} within 1\%. The second--order Bragg reflection for the radiation source Cu-K$_{\alpha1}$ and Cu-K$_{\alpha2}$ doublet at the crystal orientation (200), $2\theta\approx43^{\circ}$ is in agreement with the ratio in the diffractometer source energy within 0.02\%. Survey scans in the whole range of the diffractometer do not show additional coherent elastic scattering peaks. Nevertheless, it was not possible to identify the small peak in the position $2\theta\approx41^{\circ}$ but this could be related to SiO$_{2}$ or other impurities in the MgO crystal~\cite{Harris:2013}. 

Mg$^{2+}$ and O$^{2-}$ ions connected by an ionic bond in the MgO crystal lattice change their oscillation frequency as a function of an external perturbation energy. This external perturbation modifies the unperturbed Hamiltonian of the crystal lattice and therefore the eigenvalues and eigenfunctions associated with the oscillation frequencies of the Mg$^{2+}$ and O$^{2-}$ ions above their equilibrium point. As a consequence, when one measures the variation of a single diffracted spot by means of LEED as a function of a controlled thermal perturbation (I-T curves, as described in Section~\ref{section:video}), it is possible to obtain physical information about the stiffness of the crystal lattice. As a rule, the Debye theory of solids, which considers a crystalline solid as an elastic continuum with the thermal vibrations presented by a set of standing waves, i.e. phonons, is used to determine the Debye temperature. Generally speaking, the Debye temperature is a parameter of a solid that establishes the temperature dependence (external perturbation) of the heat capacity of the solid at constant volume~\cite{Debye:1912}.

\begin{figure}[H]
\begin{center}
\includegraphics[height=5cm]{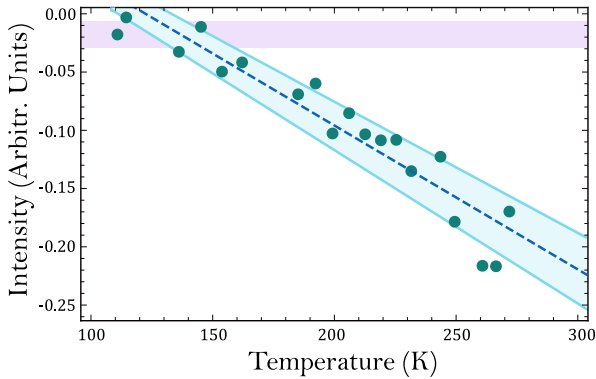}
\caption{(Color online) Temperature dependence of the intensity of a single diffracted spot. The energy of the incident electrons is 207\,eV. For a guide to the eye, a natural logarithm of the intensity for each individual diffracted spot was applied. Each data point corresponds to a video of 10\,s in which an average of around 150 measurements were performed. The blue light area is the standard deviation of a linear fit to the data. The background purple upper curve corresponds to a random area in the phosphorous screen analyzed outside the diffracted spot.}
\label{intensity}
\end{center}
 \end{figure}

In our experiment, MgO crystals are exposed to a change in their equilibrium positions by means of external thermal perturbations of constant heating rates. Single diffracted patterns such as the one shown in Fig.~\ref{experimental}(b) as a function of temperature are measured and later analyzed as described in Section~\ref{section:video}. Figure~\ref{intensity} shows the variation of the intensity of a single diffracted spot as a function of temperature. As expected, when the temperature of the system is increased there exists a lower probability that the eigenfunctions of the scattered electrons contribute in phase in order to coherently interact and therefore produce a brighter spot. As a consequence, the intensity of the diffracted spot reduces its brightness as a function of the temperature following the Debye--Waller factor (as shown below). The upper purple curve in Fig.~\ref{intensity} shows the background intensity as a function of temperature. The "y" axis is obtained by using the equation: $\textrm{ln}\left[\frac{I(T)-I_{background}}{I_{max}}\right]$, where $I(T)$ is the intensity of a single diffracted spot as a function of temperature, $I_{background}$ is the intensity in the vicinity of the diffracted spot and $I_{max}$ is the maximum intensity obtained in the experiment. In order to obtain accurate data for the intensity as a function of temperature, the background intensity was always subtracted. The difference in slope between the intensity of the diffracted spot and the background signal is of around three orders of magnitude, showing that indeed the external heating rate changes the physical properties of the crystal lattice. The data points in Fig.~\ref{intensity} are for the diffraction spot (10) corresponding to the O$^{2-}$ ion position in Fig.~\ref{experimental}(b). Similar curves were obtained for different beam spots with no special features.

Nevertheless, when we were performing the previous experiments, we noticed in some of the diffracted spots an abrupt change in the slope of the intensity as a function of temperature. For those specific diffracted spots, the change of slope occurs around the MgO crystal temperature of 185\,K. As reported in Ref.~\cite{Ferry:1996} using helium atom scattering and in Ref.~\cite{Heidberg:1995} using LEED, this temperature corresponds to the change of phase from the low--temperature water phase with symmetry $c(4\times2)$ to the high--temperature water phase with symmetry $p(3\times2)$. These results were confirmed theoretically in the high--temperature water phase for a water monolayer deposited on the MgO (100) surface, which shows that due to the interaction between the adsorbed molecules some of them dissociate~\cite{Giordano:1998}. It has been likewise predicted that phase transitions of the water overlayer-structures in these experiments could be related to the adsorption energies of the different symmetries which differ by no more than 13\,kJ/mol~\cite{Jug:2007}. In addition, the authors in Ref.~\cite{Rados:2011} estimated the amount of molecules per unit cell with $c(4\times2)$ symmetry for the low--temperature phase and for the high--temperature phase with symmetry $p(3\times2)$, finding ten water molecules and six water molecules, respectively. They also deduced that the high--temperature phase is the most stable, and that both structures which adsorbed water experience partial dissociation, as mentioned before.        

In our experiments, the MgO crystals have always been annealed up to around 820\,K for several hours and as a consequence the formation of magnesium hydroxide, which is normally formed through the chemical reaction:
\begin{equation}
\textrm{MgO} + \textrm{H}_{2}\textrm{O} \rightarrow \textrm{Mg}(\textrm{OH}_{2}), 
\label{equ:chemical} 
\end{equation}
could be reversed to separate the moisture present in the crystal. After the annealing, the crystal was cooled for several hours down to approximately 110\,K, leading to the adsorption of the available water in the vacuum system and therefore the formation of the low--temperature water phase. Then, the temperature of the MgO crystal was continuously increased with a constant heating rate and as a result we observed a change of phase of the water monolayers present in the MgO surface through monitoring the intensity of the diffracted spot as a function of temperature. The results shown in this paper correspond mostly to the high--temperature water phase since we took the higher slope in the plots of intensity as a function of temperature. This does not negatively impact the results of our analysis of the Debye temperature (as shown below) due to the fact that our experimental standard deviation partially covers this overestimation of the Debye temperature. However we are carrying out several experiments in order to obtain additional data in the low--temperature water phase to clearly resolve the change of phase in the monolayers above the surface of the MgO~\cite{GuevaraBertsch:2015}. This is very important because by understanding the change of phase of the water monolayers in the surface of the MgO crystals we expect that the total dipole moments of those molecules are coupled to the MgO lattice of the Mg$^{2+}$ and O$^{2-}$ ions and as a consequence change their positions in the lattice.        

\section{Surface Debye temperatures for the magnesium oxide lattice} \label{section:debye}  

In order to obtain the Debye temperature of the Mg$^{2+}$ and O$^{2-}$ ions in the MgO lattice, we use the following equation:
	\begin{equation}
	I(T)=I_{o}e^{-2M(T)},
	\label{equ:Debye} 
	\end{equation}
where $M(T)$ is the Debye--Waller factor obtained from the slope of the linear fit carried out in Fig.~\ref{intensity} and corresponds to $2M(T)=24\,m\,T\frac{\left(Ecos^2\theta+V_{o}\right)}{m_{a}k_{b}\left(\theta_{D}^{2}\right)}$, where $m$ is the mass of the electron, $E$ is the energy of the incident electrons, $\theta$ is the angle of incidence with respect to the surface normal, $V_{o}$ is the inner potential, $m_{a}$ is the mass of the surface atom, $k_{b}$ is the Boltzmann constant, and $\theta_{D}$ is the effective Debye temperature~\cite{Hove:1986}. The inner potential was taken as a constant value of 10\,eV, see Ref.~\cite{Cook:1982, Ferry:1998, Blanchard:1991}. Therefore, by using Eq.~\ref{equ:Debye} and the experimental data in Fig.~\ref{intensity}, the effective Debye temperature $\theta_{D}$ as a function of the energy of the incident electrons can be derived.  

It is well known that the surface quality of MgO crystals improves on thermal treatment and so we carried out a series of experiments with the purpose of correlating $\theta_{D}$ with annealing times and as a consequence with the grade of the surface crystal. In the experiments which are reported in this paper, MgO crystals are annealed up to around 820\,K in ultra--high vacuum atmospheres. We use three different annealing times, namely, 4, 8 and 12 hours. After the annealing procedure, the crystal was cooled down to around 110\,K and then, an uniform heating rate of 0.7\,K/min was used during the experiments which assures us an equilibrium state of the crystal and the thermal elements. 

\begin{figure}[H]
\begin{center}
\includegraphics[height=4.6cm]{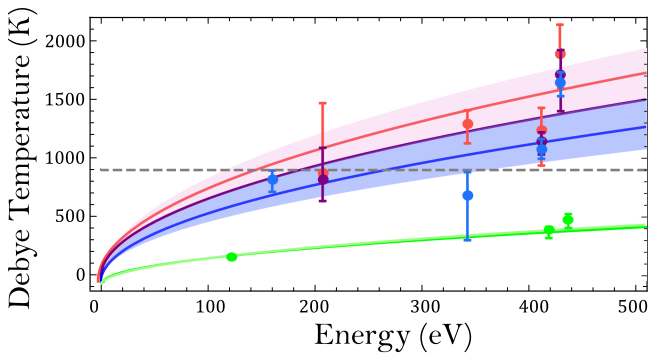}
\caption{(Color online) Effective Debye temperatures of the first three monolayers of a MgO crystal as a function of the energy of the incident electrons for three different annealing treatments. Blue light curve corresponds to 12 hours of annealing, purple curve to 8 hours and red curve to 4 hours of annealing. The horizontal line is the bulk Debye temperature reported in Ref.~\cite{Alford:2001}. Each data point is the slope of a linear fit carried out to a diffracted spot as shown in Fig.~\ref{intensity}. The error bars are obtained by means of using the maximum and minimum of the deviation in the linear fit. The green curve corresponds to the maximum intensity of a diffracted spot as was measured in Fig.~\ref{width}. See text for detailed information.}
\label{surfaceDebye}
\end{center}
 \end{figure}

Figure~\ref{surfaceDebye} shows the Debye temperature as a function of the energy for a few monolayers of the MgO crystal surface. The energy of the incident electrons could be varied from 90\,eV to 800\,eV (depending on the diffracted spots) by steps of 1\,eV but we have limited the energy range up to 440\,eV because the information coming from the top layers becomes negligible above this energy. As can be seen in Fig.~\ref{surfaceDebye}, the effective Debye temperature increases with the energy as expected from theoretical considerations. This is due to the fact that elastic penetration of the incident electrons in the crystal lattice increases as a function of the energy and as a result magnifies the multiple scattering processes inside the crystal. At the lowest energy used, i.e. 108\,eV, we derived the minimum $\theta_{D}$ which corresponds to the surface Debye temperature measured in our experiments. It was not possible to use a lower energy for the incident electrons because surface charging effects in the MgO crystal may appear and therefore the diffraction spots are unstable and ill--defined. On ther other hand, in the high energy range of the incident electrons, the $\theta_{D}$ value comes very close to the bulk value, as expected. In Fig.~\ref{surfaceDebye} the reference bulk Debye temperature is shown with a horizontal constant line~\cite{Alford:2001}. Nonetheless, in order to compare directly our experimental data of annealing (red, purple and blue curve in Fig.~\ref{surfaceDebye}) with Ref.~\cite{Alford:2001} one has to make a correction to the experimental results shown by the green curve, as shown below. In Ref.~\cite{Cook:1982} an anomalously low surface Debye temperature of 190\,K is reported. This value agrees with our result of (177 $\pm$ 10)\,K and is quite unusual since the simple rule of thumb stating that the $\theta_{D}$ value should be $\frac{1}{\sqrt{2}}$ that of the bulk~\cite{Clarke:1985} cannot be applied here because it relies on the basic assumption that the atoms in a crystal can be described as harmonic oscillators which is obviously not applicable in a system with an absence of \textit{d}-orbital electrons as the MgO lattice. 

In order to extract information of the annealing treatments related to the $\theta_{D}$ value, in Fig.~\ref{surfaceDebye} we carried out fits of the form $\theta_{D} \propto E^{1/2}$, as seen in Eq.~\ref{equ:Debye}, to compare the different annealing times. As was observed in previous experiments, we found that with higher annealing times the $\theta_{D}$ values were lower. Taking as a reference the experiment corresponding to the annealing of 8 hours, we observed that a $\Delta$t\,=\,$\pm$4\,hours produces a change of the $\theta_{D}$ value of around 15\%. We also notice that at higher annealing times, e.g. $\Delta$t\,=\,+8\,hours, the $\theta_{D}$ value starts to increase. The latter corresponds to the same observation reported in Ref.~\cite{Aswal:2002} where MgO crystals were annealed at atmospheric pressures and at different temperatures ranging from 770\,K up to 2000\,K. These authors found that annealing at temperatures higher than 1200\,K causes an increase of the MgO grains size because it reduces the surface energy of the grains, thus as a consequence they obtain rougher surfaces.

\begin{figure}[H]
\begin{center}
\includegraphics[height=5cm]{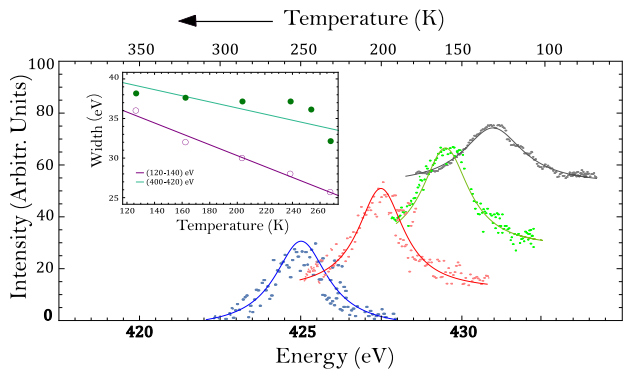}
\caption{(Color online) Intensity of a single diffracted spot as a function of the energy of the incident electrons. The inset graph shows the dependence of the FWHM (width) of a single diffracted spot as a function of the temperature (purple curve: 120--140\,eV and green: 425--432\,eV). The upper scale corresponds to the crystal temperature and grows from right to left as shown with the arrow. The solid lines are Lorentzian or linear fit (inset) to the data points. See text for further information.}
\label{width}
\end{center}
 \end{figure}

While performing the latter experiments, we noticed that the maximum intensity of a diffracted spot shifts in energy as a function of temperature. Figure~\ref{width} shows experimental evidence of the chemical shifts measured by means of LEED. As can be seen, when the total system (MgO crystal + water monolayer) is in the low--temperature water phase with symmetry $c(4\times2)$, additional energy from the incident electrons is needed. This is expected because at higher waiting times in the low--temperature water phase, higher amount of water monolayers are estimated. As we increase heating at a constant rate, we observe that the center of the maximum intensity of a diffracted spot shifts approximately --6\,eV. Our experimental observation is in good agreement with the value of the chemical shift (5.2\,eV) reported for the O--$2p$ shell in Ref~\cite{Johnson:1999} when hydroxylation of the MgO(100)-water interface is theoretically predicted. The latter motivates the argument that water can dissociate at the MgO(100)-water interface when the naturally occurring transformation of mineral periclase (MgO) to thermodynamically favored brucite (Mg(OH)$_{2}$) occurs, see Eq.~\ref{equ:chemical}. This chemical shift has also been reported using other experimental techniques, e.g. High Resolution Electron Energy Loss Spectroscopy, Ultraviolet Photoelectron Spectroscopy, X-ray Photoelectron Spectroscopy, and low--temperature Scanning Tunneling Microscopy, see Refs.~\cite{Yu:2003,Rados:2011,Shin:2010}. In our experimental results, we also observed that after adsorption of a water monolayer the entire spectrum shifts to higher electron binding energies as was explained in previous experiments~\cite{Goniakowsky:1995}. The latter is expected as a result of the positive electrostatic potential induced in the system by the presence of a proton in the surroundings of the MgO lattice. To the best of our knowledge, the measurement of this chemical shift by means of Low Electron Energy Diffraction has never been reported previously.

We carried out the same measurements described in Fig.~\ref{surfaceDebye} for an annealing time of 8\,hours (purple curve) for the MgO crystal entirely by using the maximum intensity of a diffracted spot instead of fixed parameters of the LEED electronics. The experimental results are shown in Fig.~\ref{surfaceDebye} by the green curve. The percentage difference between both curves is around 70\%. Now we can compare our experimental data with the bulk $\theta_{D}$ value from Ref.~\cite{Alford:2001}. Before doing that, it is worth mentioning that in Ref.~\cite{Alford:2001} and in other experimental observations of the $\theta_{D}$ value for the bulk, these experiments, as opposed to the experiments carried in this article, were carried out taking the crystal from higher temperatures to lower temperatures. This affects the obtained $\theta_{D}$ value since as was mentioned before there is a marked phase change from the low--temperature water phase to the high--temperature water phase. In addition, for the experiments described in Fig.~\ref{surfaceDebye} the initial conditions are different considering that we start already with few water monolayers in the surface of the MgO crystal. Therefore we carried out similar experiments (taking the MgO crystal from higher temperatures to lower temperatures) as reported in Ref.~\cite{Alford:2001} and we reach the bulk $\theta_{D}$ value of around (327 $\pm$ 30)\,eV which is in good agreement with previous measurements. We are currently undertaking measurements in order to understand the difference between having positive or negative heating rates and also studying, in a controlled way, the effects of the initial amount of water monolayers on the surface of the MgO crystals~\cite{GuevaraBertsch:2015}. We predict that applying a controlled magnetic field which controls this system that interacts in a chaotic form and simultaneously perform the positive and negative heating rates measurements we could, as in Ref.~\cite{Batal:2015}, experimentally measure the entropy production by means of coherent elastic scattering of electron diffraction.  

We also estimate the variation of the chemical shift (binding energy) due to the adsorption of water monolayers for the three annealing experiments described in Fig.~\ref{surfaceDebye}. Here, we assumed as the initial condition that the water monolayers are symmetrical and constant distributed along the MgO crystal surface. We deduce that for 4, 8 and 12 hours of annealing, the energy shifts will be 9.2, 9.0, and 8.6\,eV, respectively, as a consequence of the higher probability that the MgO surface would adsorb a water monolayer at low annealing times. We find that the water monolayers decrease their original size approximately 3\,nm when the MgO crystal is annealed for 8 hours instead of 4 hours. This is expected since at higher annealing times the MgO surface is, chemically speaking, less active and therefore one would imagine that an interaction between lattice ions and the molecules present in vacuum atmospheres is unlikely. This value is in agreement with the investigations made with atomic force microscopy and density-functional theory where for the same change in adsorption energy six water molecules in the high--temperature phase with symmetry $p(3\times2)$ were calculated~\cite{Thissen:1995}. 

In the inset of Fig.~\ref{width} we also report the full width at half maximum (FWHM) of the Lorentzian curve when we scan the energy of the incident electrons for the diffraction spot (10). The obtained width, in eV, is plotted as a function of temperature for two different penetration energies of the incident electrons. We know from diffraction that $(\frac{\lambda}{2a})^{2}=\frac{sin^2\theta}{h^2+k^2+l^2}$, where $\lambda$ is the wavelength of the incident radiation, $a$ is the lattice spacing of the cubic crystal, and $\theta$ is the scattering angle. Therefore, in our case as we have a change in the width ($\lambda$ in the previous equation) that corresponds to place water monolayers on the MgO surface assuming that the MgO lattice does not change during the experimental observations. Hence, by assuming that the MgO lattice is completely covered by symmetric and constant water monolayers, we estimate that the amount of water in the MgO surface reduces its original size approximately by 2 monolayers in the experiment where the temperature was ascending from 110\,K to 300\,K. On the other hand, in the experiment where the temperature was descending from 300\,K to 110\,K we see that the amount of water in the MgO surface increases its original size approximately by 1 monolayer. This is expected since in the former experiment we cool down the system for around 4 hours at 110\,K before performing the measurements. There is thus a higher probability that the Mg$^{2+}$ and O$^{2-}$ ions interact with the H$_{2}$O molecule and as a consequence form a water monolayer above the MgO surface. On the contrary, when the MgO surface is cooled down from 300\,K to 110\,K the probability is lower due to the fact that the surface is relatively "hot". All the above is in good agreement with the well known fact that at ultra--high vacuum atmospheres a clean surface could be covered with a complete monolayer of adsorbate in around 10$^{4}$\,s which is about our experimental time scale for the heating rate (2x10$^{4}$\,s). In addition, in the inset of Fig.~\ref{width} two different energies for the incident electrons are compared. As expected, the lower the energy of the incident electrons the higher the slope of the linear fit since the 100\,eV-electron--beam is an excellent probe for measuring the surface effects and thus is more sensitive to these changes. We also notice that if we analyze a diffracted spot corresponding to the Mg$^{2+}$ or O$^{2-}$ ions, as shown in Fig.~\ref{experimental}b), a different percentage of the water monolayers is obtained. The latter could be explained by the fact that the O$^{2-}$ ions of the H$_{2}$O molecule will interact with the Mg$^{2+}$ in the MgO lattice and as a result change their binding energy differently in the LEED pattern.                     

\section{Conclusion}

In conclusion, we have demonstrated adsorption or desorption of a few water monolayers on the magnesium oxide (100) surface lattice through a constant external thermal perturbation in the range of 110--300\,K by means of coherent elastic scattering of electron diffraction. We have additionally shown the correlation between annealing times of the surface lattice, the chemical shift of the O--$2p$ shell and the surface Debye temperature and we find that these results are in agreement with the expected phonon softening at the surface. Our experiments also determine an anomalously low surface Debye temperature around (177 $\pm$ 10)\,K, as reported in Ref.~\cite{Cook:1982}, and this sets the groundwork for a comprehensive investigation of the dynamic motion of atoms on the surface since new physical properties arise when the translational symmetry of a three-dimensional solid is broken mostly likely due to the absence of \textit{d}-orbital electrons in the MgO crystal. In order to completely resolve the change of phase by using LEED of the water monolayers on the crystal lattice from the low--temperature water phase with symmetry $c(4\times2)$ to the high--temperature phase with symmetry $p(3\times2)$, we plan to use moveable mirrors and beam splitters to collect the light from the phosphorous screen by an objective and send it either to a photomultiplier tube, to an EMCCD camera or to both detectors at the same time. This technique, usually used to increase the signal--to--noise ratio in precision spectroscopy experiments~\cite{Guggemos:2015}, can be used here in order to enhance the characteristic signal of the change of phase and also to observe other hidden phenomena in this intriguing system.     

\section{Acknowledgments}

We would like to thank V\'ictor Rodr\'iguez Araya for his expert technical support in the construction of the experiment. We also would like to thank the X-Ray Diffraction Section from the Chemistry Department of the University of Costa Rica for the XRD measurements of the MgO crystals. A special thanks to Mauricio Badilla Figueroa from the Laboratorio de Espectrometr\'ia Gamma, CICANUM, for providing us with liquid nitrogen throughout the development of these measurements. We also want to acknowledge the helpful assistance of the graphic designer Felipe Molina Guti\'errez for his effort in creating graphic art from our experimental apparatus. The authors are grateful for the support given by the Vicerrector\'ia de Investigaci\'on at the Universidad de Costa Rica to carry out this research work.

\end{document}